\documentclass[aps,prl,twocolumn,superscriptaddress,asmsymb,amsmath]{revtex4-1}
\usepackage{graphicx}
\usepackage[utf8]{inputenc}
\usepackage{hyperref}
\usepackage{blindtext}
\usepackage{braket}
\usepackage{bm}
\usepackage{siunitx}
\usepackage{color}

\begin{document}

\title{Cooper-Pair Spin Current in a Strontium Ruthenate Heterostructure}
	
\author{Suk Bum Chung}
\email{sbchung0@uos.ac.kr}
\affiliation{Department of Physics, University of Seoul, Seoul 02504, Korea}
\affiliation{Center for Correlated Electron Systems, Institute for Basic Science (IBS), Seoul National University, Seoul 08826, Korea}
\affiliation{Department of Physics and Astronomy, Seoul National University, Seoul 08826, Korea}
\author{Se Kwon Kim}
\email{evol@physics.ucla.edu}
\affiliation{Department of Physics and Astronomy, University of California, Los Angeles, California 90095, USA}
\author{Ki Hoon Lee}
\affiliation{Center for Correlated Electron Systems, Institute for Basic Science (IBS), Seoul National University, Seoul 08826, Korea}
\affiliation{Department of Physics and Astronomy, Seoul National University, Seoul 08826, Korea}
\author{Yaroslav Tserkovnyak}
\affiliation{Department of Physics and Astronomy, University of California, Los Angeles, California 90095, USA}
\begin{abstract}
It has been recognized that the condensation of spin-triplet Cooper pairs requires not only the broken gauge symmetry but also the spin ordering as well. One consequence of this is the possibility of the Cooper-pair spin current analogous to the magnon spin current in magnetic insulators, the analogy also extending to the existence of the Gilbert damping of the collective spin-triplet dynamics. The recently fabricated heterostructure of the thin film of the itinerant ferromagnet SrRuO$_3$ on the bulk Sr$_2$RuO$_4$, the best-known candidate material for the spin-triplet superconductor, offers a promising platform for generating such spin current. We will show how such heterostructure allows us to not only realize the long-range spin valve but also electrically drive the collective spin mode of the spin-triplet order parameter. Our proposal represents both a new realization of the spin superfluidity and a transport signature of the spin-triplet superconductivity.
\end{abstract}
\maketitle

\begin{figure}
	\includegraphics[width= 0.9\columnwidth]{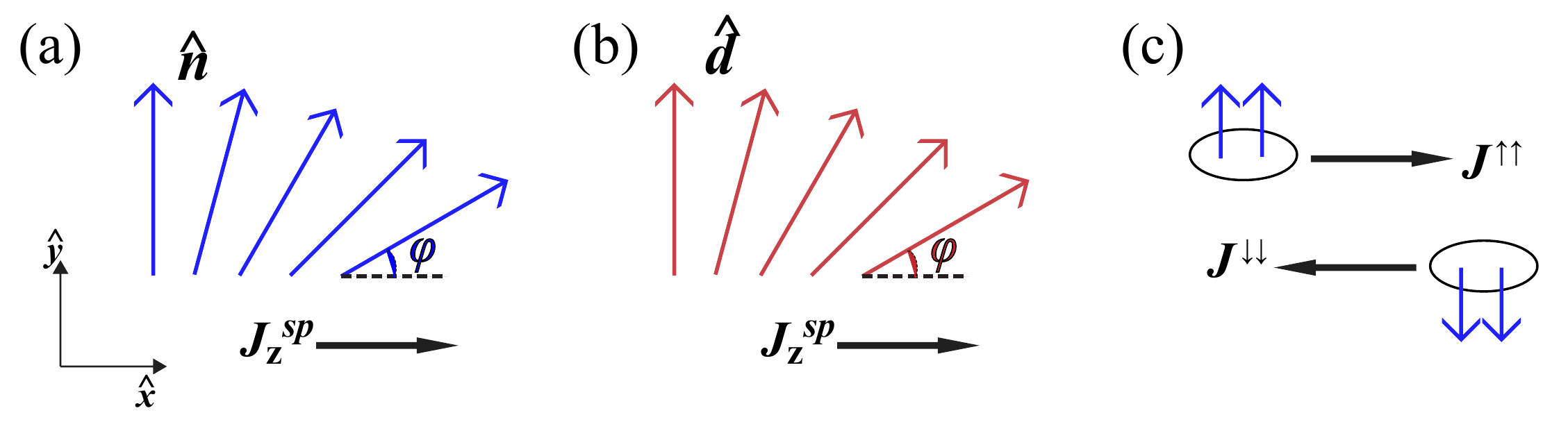}
\caption{Schematic illustration of the analogy between the magnetic insulator and the spin-triplet superconductor. (a) The planar spiraling of the magnetic order parameter ${\bf \hat{n}}$ leads to spin current. (b) The same phenomena occurs for that of the spin component ${\bf \hat{d}}$ of the spin-triplet superconductor order parameter, (c) the dual picture of which is the counterflow of the spin up-up and down-down pairs.}
\label{FIG:scheme}
\end{figure}

{\it Introduction}: Harnessing spin rather than charge in electronic devices has been a major topic in solid state physics, which not only has been utilized for various memory devices but is also expected to play a key role in processing quantum information \cite{Wolf2001}. 
In order for various spin devices to function robustly, 
the long-range spin transport needs to be achieved. Metallic wires, however, typically do not transport spins beyond the spin-diffusion length due to the 
single electron spin relaxation \cite{Bass2007}.

In recent years, it has been shown that the exponential damping can be circumvented in the spin transport via collective magnetic excitations. For example, easy-plane (ferro- and antiferro-)magnetic insulators, as the U(1) order parameter can characterize them, may be considered analogous to the conventional superfluid \cite{Takei2014L, Takei2014B, Chen2014}. As Fig.~\ref{FIG:scheme} (a) illustrates schematically, the planar spiraling of the magnetic order parameter in such magnetic insulators can give rise to the spin supercurrent, just as the phase gradient of the conventional superfluid gives rise to the mass supercurrent; in this sense these magnetic insulators can be regarded as spin superfluids \cite{Sonin}.

Interestingly, there exists a class of superfluids and superconductors which can support both mass and spin supercurrent. Such superfluids and superconductors would need to involve both spin ordering and gauge symmetry breaking. 
This occurs in the condensate of both the spin-1 bosons \cite{StamperKurn2013} and the spin-triplet Cooper pairs of $^3$He atoms \cite{Leggett1975, Vollhardt1990} or electrons \cite{Sigrist1991, Mackenzie2003}; in the latter case, the dissipationless spin current would be carried by the Cooper pairs. While 
the vortices with spin supercurrent circulation have been observed in all theses systems \cite{Jang2011, Seo-Autti}, the bulk spin supercurrent 
has not been detected 
in the superconductor.

In this Letter, we will show how this existence of spin superfluidity in the spin-triplet superconductor allows not only the long-range spin current but also electrically exciting the spin wave in the bulk. For realizing these phenomena, we propose a two-terminal setup with voltage bias between ferromagnetic metal leads 
in contact with the spin-triplet superconductor. 
While the static order-parameter case \cite{Romeo2013} can be essentially reduced to the Blonder-Tinkham-Klapwijk type formalism \cite{Blonder1982} for the interfacial transport, here we need to complement it with the appropriate equations of motion for the collective spin dynamics in the superconductor.
Recently, a thin film of the itinerant ferromagnet SrRuO$_3$ 
has been epitaxially deposited on 
the 
bulk Sr$_2$RuO$_4$, the best known candidate material for the spin-triplet superconductor 
\cite{tripletSRO}, yielding, due to their structural compatibility, an atomically smooth and highly conductive interface \cite{Anwar2015}  
with a strong 
Andreev conductance 
\cite{Anwar2016}. 
This makes Sr$_2$RuO$_4$ and SrRuO$_3$ 
the most suitable candidate materials for the bulk and the leads, respectively, of our setup \footnote{LCMO is one possible substitute for SrRuO$_3$.}. 
For the remainder of this paper, we will first show how the simplest effective spin Hamiltonian for the spin-triplet superconductor and the resulting spin dynamics are analogous to those of the antiferromagnetic insulator; then, we will discuss the magnetoresistance for the DC bias voltage 
and the coupling between the AC bias voltage and the spin wave.

{\it General considerations}: We first point out the close analogy between the spin order parameter of the antiferromagnet and the spin-triplet superconductor. Defined 
\begin{equation}
i({\bf d}\cdot{\bm \sigma})\sigma_y \!=\! \left[\begin{array}{cc} -d_x + i d_y & d_z\\
                                                                                                                d_z & d_x + i d_y \end{array}\right]
                                                         \!\equiv\! \left[\begin{array}{cc} \Delta_{\uparrow\uparrow} & \Delta_{\uparrow\downarrow}\\
                                                                                                           \Delta_{\downarrow\uparrow} & \Delta_{\downarrow\downarrow}\end{array}\right],
\label{EQ:tripletOP}                                                                                                     
\end{equation}
the ${\bf d}$-vector of the spin-triplet pairing, which parametrizes the Cooper-pair spin state, behaves similarly under spin rotations to the N\'eel order parameter of an antiferromagnet, i.e., $[S_i ({\bf r}), d_j ({\bf r}')] = i \hbar\epsilon_{ijk}\delta({\bf r}-{\bf r}') d_k ({\bf r})$ and $[d_i, d_j] = 0$ for the condensate spin ${\bf S}$ 
(unlike the magnetization, neither the N\'eel order parameter nor the ${\bf d}$-vector generate the spin rotation in themselves) \cite{Leggett1975, Vollhardt1990, Mackenzie2003}. Given that the commutation relations establish ${\bf S}\times {\bf \hat{d}}$ as the conjugate momentum to ${\bf d}$ in both cases, it is natural that the simplest effective  Hamiltonian for the spin-triplet superconductor ${\bf \hat{d}}$-vector, 
\begin{equation}
H = \frac{1}{2} \int d{\bf r} [A (\nabla {\bf \hat{d}})^2 + K\hat{d}_z^2 + \gamma_e^2{\bf S}^2/\chi],
\label{EQ:Ham}
\end{equation}
where $\gamma_e$ is the electron gyromagnetic ratio, $A$ the ${\bf \hat{d}}$-vector stiffness, and $\chi$ the magnetic susceptibility, should be equivalent to that of the antiferromagnet N\'eel order parameter, 
once we identify the ${\bf \hat{d}}$-vector with the N\'eel order parameter \cite{Takei2014B}. In the latter, antiferromagnetic case, a ($xy$) planar texture of the orientational order parameter $\mathbf{\hat{n}}\to(\cos\phi,\sin\phi,0)$ is associated with a collective ($z$-polarized) spin current $J_z\propto\mathbf{z}\cdot\mathbf{\hat{n}}\times\partial_i\mathbf{\hat{n}}\to\partial_i\phi$ flowing in the $i$th direction.
While this extends directly to our spin-triplet case, Eq.~\eqref{EQ:tripletOP} gives 
the intuitive dual picture of Fig.~\ref{FIG:scheme} (c) 
for the 
planar spiraling of the ${\bf d}$-vector, 
i.e., ${\bf \hat{d}} = (\cos \alpha, \sin \alpha, 0)$. Namely, as the phase of 
$\Delta_{\uparrow\uparrow}$ ($\Delta_{\downarrow\downarrow}$) 
is given by $\phi_c \mp \alpha$ (where $\phi_c$ is the overall phase of the superconductor), the spiraling of the ${\bf d}$-vector on the $xy$ plane as shown in Fig.~\ref{FIG:scheme} (b), or the gradient of $\alpha$, would imply the counterflow of the spin up-up and down-down pairs. The resultant ($z$-polarized) spin current is $\propto -\bm{\nabla}\alpha$. Given the same commutation relation and the same effective Hamiltonian, it is natural that, in absence of dissipation, the equations of motion for these two cases, 
the Leggett equations 
the ${\bf \hat{d}}$-vector \cite{Leggett1974, Leggett1975, Vollhardt1990} and the Landau-Lifshitz type equation for the N\'eel order parameter, are identical.

We further argue that both cases have the same phenomenological form of dissipation as well. For the case of the N\'eel order parameter $\mathbf{\hat{n}}$, such dissipation, $\propto\alpha(\partial_t\mathbf{\hat{n}})^2$, known generally as Gilbert damping for collective magnetic dynamics, has been understood phenomenologically 
\cite{Hals2011, Takei2014B, Kim2014}. That such dissipation has not been featured in the 3He superfluid literature can be attributed not to the intrinsic nature of the spin-triplet pairing but rather to the very weak relativistic spin-orbit coupling of the 3He atoms originating solely from the nuclear dipole-dipole interaction \cite{Leggett1975}. In contrast, electrons in Sr$_2$RuO$_4$ are subject to the Ru atomic spin-orbit coupling \footnote{Given the effect of the chemical potential on the magnitude of the orbital hybridization induced by the Ru spin-orbit coupling at the Fermi surface, a strong electron density nonuniformity may affect the spin relaxation of the Cooper-pair condensate.} estimated to be of order 0.1~eV \cite{Haverkort2008}. 
In this work, we will consider the decay rate of $\alpha n \hbar  \gamma_e^2/\chi$ for the condensate spin, the addition of which makes the Leggett equations of motion for spin \footnote{Following \cite{Chung2007}, we leave out in this work for simplicity any possible complications in Sr$_2$RuO$_4$ due to the orbital degrees of freedom \cite{Chung2012, Sauls2015, Huang2016}.} equivalent to the Landau-Lifshitz-Gilbert type equations for antiferromagnets:
\begin{align}
\partial_t  {\bf \hat{d}} =& -{\bf \hat{d}} \times \frac{\gamma_e^2}{\chi}{\bf S},\nonumber\\
\partial_t {\bf S} =& {\bf \hat{d}} \times (A\nabla^2 {\bf \hat{d}}-K \hat{d}_z {\bf \hat{z}} - \alpha n\hbar\partial_t {\bf \hat{d}}),
\end{align}
where $\alpha$ is the dimensionless Gilbert damping parameter and $n$ the Cooper-pair density. This set of equations shows how the effective Hamiltonian of Eq.~\eqref{EQ:Ham} provides the simplest method for considering the local ${\bf \hat{d}}$-vector dynamics, including the spin-wave excitation and the collective dissipation.

For the boundary conditions, at the interface between the ferromagnetic lead and the spin-triplet superconductor, we consider a two-channel interface conductance due to the spins aligned or anti-aligned to the lead magnetization
We note, in this regard, that the SrRuO$_3$ thin film has a very high transport spin polarization, with a 3-to-1 ratio between the majority and minority spin channels \cite{Nadgorny2003, Raychaudhuri2003, Shiga2017}, while the magnetization gets enhanced in the heterostructure \cite{Anwar2015}.
In this Letter, for the sake of simplicity, we shall only consider the case where the lead magnetizations are collinear. Furthermore, the ${\bf d}$-vector of the bulk spin-triplet superconductor will be taken to be perpendicular to the lead magnetization, i.e., the Cooper pairs are equal-spin paired along the quantization axis parallel to the magnetization; it has been claimed for the Sr$_2$RuO$_4$ superconductor, based on the $c$-axis NMR measurement, that its ${\bf d}$-vector can be rotated into the $ab$-plane by applying magnetic field larger than 200~G \cite{Murakawa2004}, well below the upper critical field.

\begin{figure}
	\includegraphics[width= 0.8\columnwidth]{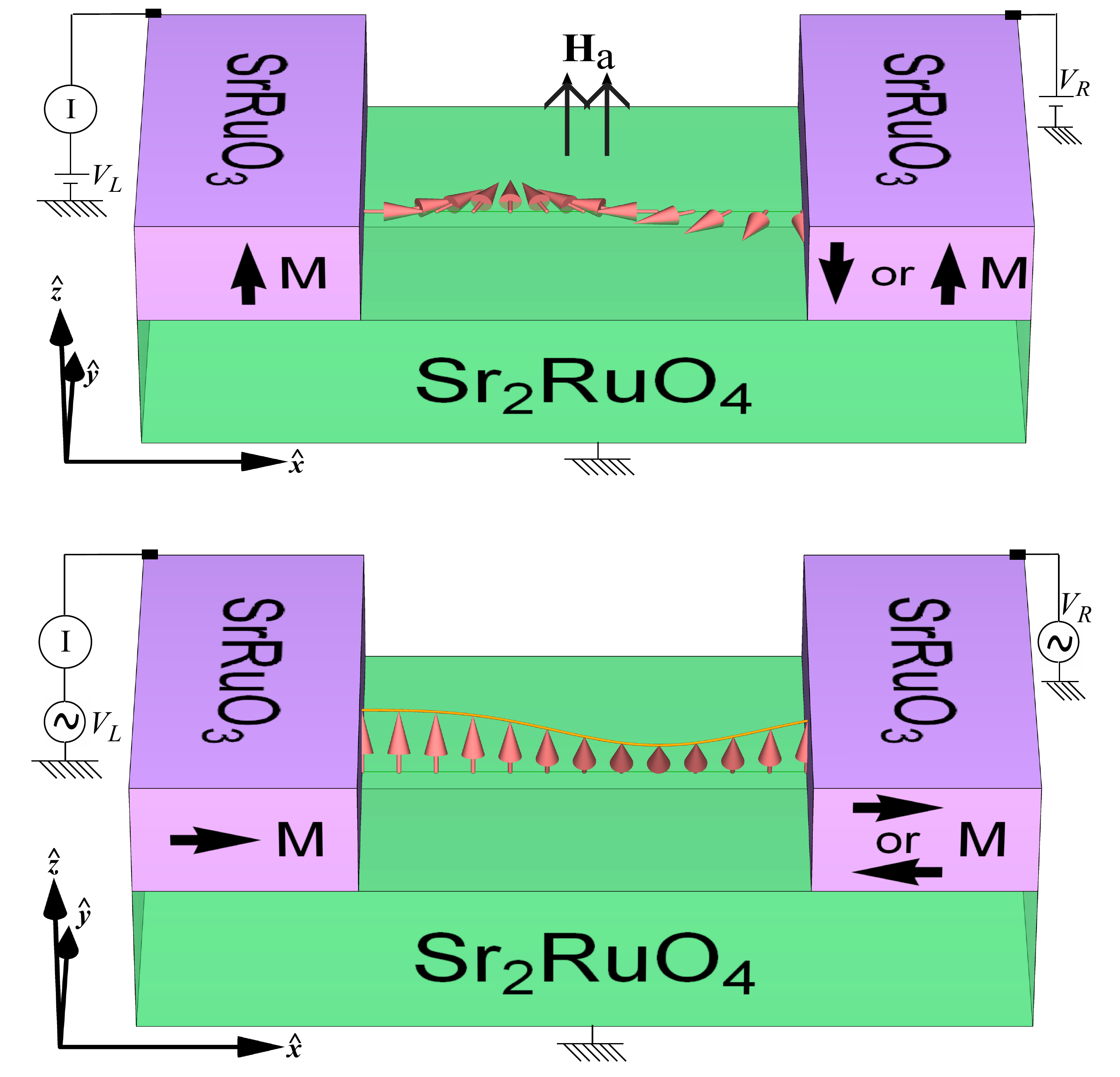}
\caption{The setup for the DC voltage bias for the spin valve (upper) and the AC bias voltage for the spin-wave detection (lower),
where ${\bf \hat{x}},{\bf \hat{y}},{\bf \hat{z}}$ coincide with the crystalline $a,b,c$-axes, respectively. 
For the upper figure, the lead magnetization is along 
the $c$-axis, with the applied magnetic field $H_a \geq 200$~G along the $c$-axis giving us the easy plane ${\bf d}$-vector configuration on the $ab$-plane, hence the 
spiraling in the $ab$-plane. For the lower figure, the lead magnetization is along 
the $a$-axis; as the easy-axis ${\bf d}$-vector anisotropy favors the alignment along the $c$-axis, in the absence of an applied filed, the AC bias voltage gives us the low-frequency standing wave of the ${\bf d}$-vector oscillating around the $c$-axis in the $bc$-plane.}
\label{FIG:setup}
\end{figure}

{\it Long-range spin valve}: The simplest physics that can arise in our two-terminal setup is the spin-valve magnetoresistance due to the relative alignment of the leads. We consider the case where the spin-triplet superconductor has the easy-plane anisotropy, that is, $K>0$ in Eq.~\eqref{EQ:Ham}, while the lead magnetization is perpendicular to this plane; as already mentioned, the former can be realized for the SrRuO$_3$/Sr$_2$RuO$_4$ heterostructure by applying a $\geq 200$~G field along the $c$-axis. In this case, we can take $\hat{d}_z$ to be a small parameter in ${\bf \hat{d}} = (\sqrt{1-\hat{d}_z^2}\cos \phi_z, \sqrt{1-\hat{d}_z^2}\sin \phi_z, \hat{d}_z)$ and $|S_{x,y}| \ll |S_{z}|$. In such a case, $[\phi_z ({\bf r}), S_z ({\bf r}')] = i\hbar \delta({\bf r}-{\bf r}')$ gives us the conjugate pair,
leading to the equations of motion
\begin{equation}
\partial_t \phi_z = \frac{\gamma_e^2}{\chi} S_z,\,\,\,\,\,\,\,\,\partial_t S_z = A \nabla^2 \phi_z - \alpha n \hbar  \partial_t \phi_z,
\label{EQ:spinSink}
\end{equation}
where the first equation is a spin analogue of the Josephson relation and the second is the spin continuity equation with the relaxation term. Note that we measure $S_z$ with respect to its equilibrium value. One confirms the condensate spin imbalance relaxation time to be $\chi/\alpha n \hbar \gamma_e^2$ from Eq.~\eqref{EQ:spinSink} through  
deriving $\partial_t S_z + \boldsymbol{\nabla} \cdot {\bf J}^{sp}_z = - \alpha n \hbar  \gamma_e^2 S_z/\chi $,
where ${\bf J}^{sp}_z = -A \boldsymbol{\nabla} \phi_z$. 
It is also important to note here that the magnitude of the $d$-vector anisotropy $K$ has no effect on the 
in-plane $d$-vector precession, 
which allows us to ignore the fact that our applied field gives us the Abrisokov vortices in the spin-triplet superconductor and hence a non-uniform $K$.

We consider the spin-up current and the spin-down current to be independent at the interface:
\begin{equation}
I^\sigma_{L,R} = \pm g^{\sigma\sigma}_{L,R} (V_{L,R} - \hbar\partial_t \varphi_\sigma/2e), 
\label{EQ:BC}
\end{equation}
where $g^{\sigma\sigma}_{L,R}$'s are the conductances for the $\sigma$-spin, $I_{L,R}$ the $\sigma$-spin current into (out of) the left (right) lead, and $V_{L,R}$ the bias voltage of the left (right) lead; this is due to the spin-triplet superconductor having the equal spin pairing axis collinear with the lead magnetization and taking $g^{\uparrow\downarrow} = 0$. 
From Eq.~\eqref{EQ:tripletOP}, we see that the overall (or charge) phase of the superconductor is given by the average of the spin up-up and the spin down-down condensate phase, $\phi_c = \sum_\sigma  \varphi_\sigma/2$, while $\phi_z$ of Eq.~\eqref{EQ:spinSink} is given by $\phi_z = \sum_\sigma \sigma \varphi_\sigma/2$. We are interested here in the steady-state solution, i.e., $\partial_t \varphi_\sigma = {\rm const}$, for which we define the constant precession rate of  $\omega_c \equiv \sum_\sigma \partial_t \varphi_\sigma/2$ for the overall phase $\phi_c$ and  $\Omega_s \equiv \sum_\sigma \sigma \partial_t \varphi_\sigma/2$ for $\phi_z$. For such solution, the following continuity conditions can be applied to the charge and spin supercurrents, respectively:
\begin{equation}
\sum_\sigma (I^\sigma_L - I^\sigma_R) \!=\! 0,\,\,\,\,\,\,\,\sum_\sigma \sigma(I^\sigma_L - I^\sigma_R) \!=\! 2\alpha  n e  \Omega_s SL
\label{EQ:continuity}
\end{equation}
($S$ is the bulk cross section area and $L$ the spacing between the two leads), the former from the charge conservation and the latter from applying the steady-state condition on Eq.~\eqref{EQ:spinSink}, along with the spin current loss $\propto\alpha L$ in the superconductior. 

The current through the Sr$_2$RuO$_4$ bulk can be obtained from the interface boundary conditions and the continuity conditions above, with the larger magnitude for the parallel magnetization than the antiparallel magnetization. We define the total conductance 
$g_{L,R} \equiv \sum_\sigma g^{\sigma\sigma}_{L,R}$ and the conductance polarization $p_{L,R} \equiv \sum_\sigma \sigma g^{\sigma\sigma}_{L,R}/g_{L,R}$, which defines the relevant transport spin polarization. 
Applying the continuity conditions Eq.~\eqref{EQ:continuity} on the interface boundary conditions Eq.~\eqref{EQ:BC} and setting $V_L = -V_R = V/2$, 
we obtain 
\begin{equation}
\left(\!\begin{array}{cc} g_L\!+\!g_R & p_L g_L\!+\!p_R g_R\\ p_L g_L\!+\!p_R g_R & g_L\!+\!g_R\!+\!g_\alpha \end{array}\!\right)\!\left(\!\begin{array}{c} \omega_c \\ \Omega_s \end{array}\!\right)  
\!=\! \frac{eV}{\hbar}\!\left(\!\begin{array}{c} g_L\!-\!g_R \\ p_L g_L\!-\!p_R g_R \end{array}\!\right),
\end{equation}
where $g_\alpha \equiv \frac{4\alpha n e^2 SL}{\hbar}$. 
We can now obtain 
the dependence of the charge current 
on the conductance polarization:
\begin{equation}
I^c 
\!=\!\sum_\sigma I^\sigma \!=\! I_0\!\left[\!1\!-\!\frac{g_L g_R (p_L \!-\! p_R)^2}{(g_L \!+\! g_R) (g_L \!+\! g_R \!+\! g_\alpha) - (p_L g_L \!+\! p_R g_R)^2}\!\right],
\label{EQ:MCurrent}
\end{equation}
where $I_0 \equiv g_L g_R V/(g_L + g_R)$. Note that $I^c$ is maximized at $p_L = p_R$, when the steady-state angle $\phi_z$ remains static. Different spin polarizations at the two ends, on the other hand, would trigger spin dynamics and result in a nonzero dissipation rate of  $R = \frac{1}{2} \alpha n \hbar \Omega_s^2 = 
R_0 (1-I^c/I_0)^2 / (p_L - p_R)^2$ per volume of the superconducting bulk, where $R_0=8\alpha n (eV)^2/\hbar$. 
Given that 
$p_{L,R}$ 
change sign on the magnetization reversal, 
the above results effectively give us the spin-valve magnetoresistance of our 
heterostructure, i.e., a larger conductance for the parallel magnetizations than for the antiparallel. Any effect that the spin-triplet pairing may have on the magnetization, hence the conductance polarization, can be ignored when the 
Curie temperature of SrRuO$_3$ ($\sim$ 160K) \cite{Koster2012} is two orders of 
magnitude higher than the superconducting critical temperature ($\sim$ 1.5K) Sr$_2$RuO$_4$.

We emphasize that the above magnetoresistance result is obtain solely for the current carried by Cooper pairs. 
At a finite-temperature, quasiparticle contribution would generally result in an exponentially-decaying magnetoresistance, negligible for the lead spacing beyond the spin-diffusion length. By contrast, the current of Eq.~\eqref{EQ:MCurrent}, which is carried by the Cooper pairs, gives us the $\sim 1/L$ behavior for the large spacing limit. Therefore, any magnetoresistance beyond the quasiparticle spin-diffusion length should arise only below the superconducting transition at $T_c$, upon the emergence of a Cooper-pair condensate. For our Sr$_2$RuO$_4$ / SrRuO$_3$ heterostructure, detection of magnetoresistance in the superconducting state for the lead spacing larger than the Sr$_2$RuO$_4$ spin-diffusion length
can be taken as a transport evidence for the spin-triplet superconductivity. The value of the spin-diffusion length itself can be extracted by measuring the exponential decay of the (normal) magnetoresistance, both above and below the transition.

\begin{figure}
	\includegraphics[width= 0.8\columnwidth]{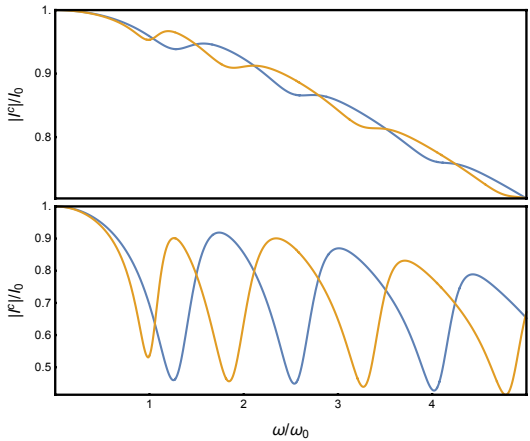}
\caption{Charge current versus frequency plotted for $\tilde{g} = 0.5$, $\tilde{L} = 2$, $\Gamma/\omega_0 = 0.1$ and $\tilde{A} = 0.2$, with the orange curve representing $p_L = p_R = p$ and the blue $p_L = -p_R = p$. 
Note that $p = 0.8$ for the top plot and $p=0.2$ for the bottom plot.}
\label{FIG:biasAC}
\end{figure}

{\it Electrically driven spin collective mode}: For the case of the easy-axis anisotropy of the ${\bf d}$-vector, hence $K<0$ in Eq.~\eqref{EQ:Ham},the spin collective excitation of the Cooper pairs \cite{Leggett1975, Vollhardt1990, collectiveSRO, Chung2012} will modify the supercurrent transport under the AC bias voltage. 
We shall still continue to consider the case where Eq.~\eqref{EQ:BC} 
would be valid, i.e., the equal spin pairing axis of the spin-triplet superconductor collinear to the lead magnetizations. One way to satisfy this condition would be to have the lead magnetizations collinear to the $a$-axis, with no applied magnetic field; that would leave the $a$-axis as the equal spin pairing axis, with 
the ${\bf d}$-vector moving on the the $bc$-plane. The equations of motion, corresponding to spin injection polarized along the $x$-direction, are then modified to
\begin{equation}
\partial_t \phi_x \!=\! \frac{\gamma_e^2}{\chi} S_x,\,\,\,\,\,\,\partial_t S_x \!=\! A \nabla^2 \phi_x \!-\! \omega_0^2 \frac{\chi}{\gamma_e^2}\!\cos \phi_x \!\sin \phi_x \!-\! \alpha \hbar \partial_t \phi_x,
\label{EQ:spinSink2}
\end{equation}
where $\phi_x$ is conjugate to $S_x$ and $\omega_0^2 \equiv |K|\gamma_e^2/\chi$ is the spin-wave energy gap. 
For the AC voltage bias $V=V_0 \exp(-i\omega t)$, the steady-state solution for the spin phase $\phi_x (x,t) = f(x) \exp(-i\omega t)$ and the charge phase $\phi_c (x,t) = g(x) \exp(-i\omega t)$ behave differently, focusing on the frequencies far below the plasma frequency. Hence 
the spin equations of motion Eq.~\eqref{EQ:spinSink2} gives us $f(x) = C_+ \cosh \kappa x + C_- \sinh \kappa x$, where $v^2\kappa^2 = \omega^2 - \omega^2_0 -i\omega \Gamma$, with $v \equiv \gamma_e \sqrt{A/\chi}$ (the ${\bf \hat{d}}$-vector stiffness $A$ defined in Eq.~\eqref{EQ:Ham}) being the spin-wave velocity and $\Gamma \equiv \alpha n \hbar \gamma_e^2/\chi$ the damping rate. 
By contrast, the charge current $J^c (x,t) =  -\rho \partial_x \phi_c$, where $\rho$ is the $\phi_c$ stiffness, 
should be uniform, 
which means we can set $\phi_c (x,t) = {\rm const.} -x (J^c_0/\rho)  \exp(-i\omega t)$, with a constant $J^c_0$. By imposing consistency between the current obtained from the boundary conditions of Eq.~\eqref{EQ:BC} and the dynamics of Eq.~\eqref{EQ:spinSink2}, we can solve for 
$J^c_0$ and $C_\pm$;  Fig.~\ref{FIG:biasAC} shows the numerical results for $I^c = J^c_0 S$ for the case of both $p_L = p_R$ and $p_L = -p_R$. 

Our numerical results 
show that magnetoresistance becomes significant at $\omega \gtrsim \omega_0$, where the collective spin mode of the Cooper pairs is activated. 
For simplicity we have set $g_L = g_R = g$ and used the dimensionless parameters $\tilde{g} \equiv g \hbar v/2eA$, $\tilde{L} \equiv \omega_0 L/2v$, and $\tilde{A} = A/\rho$. For $\omega < \omega_0$, in addition to 
barely noticeable magnetoresistance, the charge current amplitude does not oscillate with frequency; 
it remains close to the DC value $I_0$, which contrasts with the complete transport suppression obtained for the magnetic insulator \cite{Takei2014L}.
In contrast, for $\omega > \omega_0$, we see an oscillation with the $\omega/\omega_0$ period of about $\pi/\tilde{L}$, where 
the current amplitude maxima for the antiparallel lead magnetization occur at the current amplitude minima for the parallel lead magnetization and vice versa. As in the ferromagnetic insulator \cite{Takei2014L}, 
we expect that for $\tilde{L} \ll 1$ (while $L$ is still larger than the quasiparticle spin-diffusion length), the magnetoresistance of Eq.~\eqref{EQ:MCurrent} is recovered for the static bias, {\it i.e.}, $\omega \to 0$. 

We point out that the detection of the oscillation shown in Fig.~\ref{FIG:biasAC} would determine the yet-unknown energy parameters for the spin-triplet pairing of Sr$_2$RuO$_4$. From the effective Hamiltonian of Eq.~\eqref{EQ:Ham}, if we had known accurately the field $H_c$ along the $c$-axis that would exactly restore the ${\bf d}$-vector isotropy, the gap frequency $\omega_0$ should be just the electron Larmor frequency of this field from the spin equations of motion of Eq.~\eqref{EQ:spinSink2}. However, we know no more than the upper bound $H_c < 200$~G, hence only $\omega_0 < \gamma_e \times 200~{\rm G} = 3.5$~GHz, 
while the AC bias experiment, as shown in in Fig.~\ref{FIG:biasAC}, would allow us to definitely identify the spin collective mode gap.

{\it Conclusion and discussion}: We have studied the DC and AC current transport between the itinerant ferromagnetic lead with collinear magnetization through the spin-triplet superconductor. We showed here that magnetoresistance can arise for both cases due to the Cooper-pair spin transport. For the DC bias, the persistence of magnetoresistance for the lead spacing larger than the quasiparticle spin-diffusion length can be taken as a transport evidence for the spin-triplet pairing. For the AC bias, the activation of magnetoresistance and frequency dependent oscillation above the threshold frequency will allow us to determine the spin anisotropy energy scale. All together, our work shows both a new realization of the spin superfluidity and a transport signature of the spin-triplet superconductivity. The recently fabricated SrRuO$_3$/Sr$_2$RuO$_4$ heterostructure provides a promising experimental setup.

{\it Acknowledgement}: We would like to thank Young Jun Chang, Bongju Kim, Han Gyeol Lee, Seung Ran Lee, Yoshiteru Maeno, Tae Won Noh, S. Raghu, Manfred Sigrist and So Takei for sharing their insights. The hospitalities of Natal International Institute for Physics during the workshop ``Collective Spin Transport in Electrical Insulators" (S.B.C. and Y.T.) and of Kyoto University during the workshop ``Oxide Superspin 2017" (S.B.C.), where parts of this work has been completed, are gratefully acknowledged. This research was supported by IBS-R009-Y1 (S.B.C. and K.H.L.), 
and United States Army Research Office under Contract No. W911NF-14-1-0016 (S.K.K. and Y.T.).

\bibliography{ref_SRO-spin}

\end{document}